\documentclass[pre,showpacs,twocolumn,longbibliography,nofootinbib,twoside]{revtex4-1}

\usepackage[colorlinks,allcolors=blue]{hyperref}
\usepackage{graphicx}
\usepackage{amsmath,mathrsfs}
\usepackage{amsfonts}
\usepackage{amssymb}
\usepackage{wasysym}
\usepackage{dsfont}
\usepackage{color,soul}

\begin{document}

\title{Stochastic Kuramoto oscillators with discrete phase states}

\author{David J. J\"org}\email{djj35@cam.ac.uk}
\affiliation{Theory of Condensed Matter Group, Cavendish Laboratory, University of Cambridge, JJ Thomson Avenue, Cambridge CB3 0HE, United Kingdom}
\affiliation{The Wellcome Trust/Cancer Research UK Gurdon Institute, University of Cambridge, Tennis Court Road, Cambridge CB2 1QN, United Kingdom}

\date{\today}

\begin{abstract}
\noindent We present a generalization of the Kuramoto phase oscillator model in which phases advance in discrete phase increments through Poisson processes, rendering both intrinsic oscillations and coupling inherently stochastic. We study the effects of phase discretization on the synchronization and precision properties of the coupled system both analytically and numerically. Remarkably, many key observables such as the steady-state synchrony and the quality of oscillations show distinct extrema while converging to the classical Kuramoto model in the limit of a continuous phase. The phase-discretized model provides a general framework for coupled oscillations in a Markov chain setting.
\end{abstract}


\maketitle

\noindent  

\section{Introduction}

\noindent%
The dialectic of synchronization has become a powerful conceptual tool in theoretical physics---rooted in the description of coupled oscillators and clocks \cite{Strogatz1993}, it has been extended to phenomena that bear only structural resemblance to coupled oscillators such as the collective behavior of bird flocks \cite{Ha2010} and magnetic systems \cite{Flovik2016}.
Hence, it is not surprising that probably the most prominent theoretical paradigm for synchronization, the celebrated Kuramoto model of coupled phase oscillators and its multifarious variants \cite{Kuramoto1984,KuramotoBook1984,Acebron2005,Rodrigues2016}, have been applied to problems as different as neutrino oscillations \cite{Pantaleone1998}, embryonic body axis segmentation \cite{Giudicelli2007,Morelli2009,Jorg2016}, electric power grids \cite{Filatrella2008,Rohden2012,Dorfler2012}, epileptic seizures \cite{Schmidt2014}, and quantum entanglement \cite{Witthaut2017}.
The Kuramoto model is a time-continuous, phase-continuous system of coupled differential equations \cite{KuramotoBook1984,Acebron2005,Rodrigues2016},
\begin{align}
	\frac{\mathrm{d}\phi_i}{\mathrm{d}t} = \omega_i + \kappa \sum_{j=1}^N c_{ij} \Gamma(\phi_j-\phi_i) \ , \label{eq.kuramoto}
\end{align}
where $\phi_i$ is the phase of oscillator $i=1,\hdots,N$ and $\omega_i$ is its intrinsic frequency, $\kappa$ is the coupling strength, $\Gamma$ is a $2\pi$-periodic coupling function, and $c_{ij}$ is the coupling topology matrix where $c_{ij} > 0$ indicates that the dynamics of oscillator $i$ couples to the dynamics of oscillator~$j$ and $c_{ij}=0$ otherwise.
For appropriate choices of $\Gamma$ and $c_{ij}$, the coupling term alters the dynamic frequency $\mathrm{d}\phi_i/\mathrm{d}t$ in such a way that the system tends to synchronize, given that coupling can overcome the spread in frequencies \cite{Strogatz2000}.

Whether a phase-continuous model is a viable description depends on the system at hand.
Biochemical oscillators, for instance, operate through chemical and/or genetic feedbacks between different molecule species and are often characterized by small numbers of molecules which are subject to fluctuations \cite{Morelli2007,Zwicker2010,Suvak2012,Webb2016,Lengyel2017}.
Another prominent example from biology is the cell cycle, which, while going through well-defined states, can exhibit considerable period variations \cite{Weber2014}.
Often, it is desirable to represent such processes on a coarse-grained level, e.g., by Markov chain models, when only their core features are to be retained.
This is especially interesting if coupled oscillatory processes are part of a more complex system involving interactions with non-oscillatory parts. The latter is often the case in biology, where periodic processes interact with cell fates, intercellular signaling systems, and/or tissue growth \cite{Sugimoto2012,Brown2014,Jorg2016}.
In recent years, there has been an extensive interest in the behavior of discrete-state models of uncoupled and coupled oscillators \cite{Prager2003,Wood2006,Wood2007,Fernandez2008,Tonjes2011,Suvak2012,Pinto2014,Escaff2014,Barato2016}.
Recently, for instance, the question has been investigated whether discrete-state models can capture the behavior of the noisy Kuramoto model with all-to-all coupling and homogeneous frequencies \cite{Escaff2016}.

In this paper, we study a generalization of the Kuramoto model in which each oscillator transitions between discrete phase states with defined transition rates.
This renders all parts of the model inherently stochastic, including the coupling dynamics between oscillators.
We investigate the effects of phase discretization on the dynamics of systems with homogeneous and inhomogeneous frequencies, in particular their synchronization behavior, their phase-coherence, and their period fluctuations.
In Section~\ref{sec:single.osc}, we introduce the description of a single phase-discretized oscillator and characterize its stochastic properties such as its effective frequency and its quality factor.
In Section~\ref{sec:osc.coupling}, we introduce a stochastic generalization of the coupled Kuramoto model with arbitrary coupling topology and coupling function and discuss variants of this generalization.
In Section~\ref{sec:two.oscillators}, we study the case of two coupled oscillators and investigate the effects of coupling on synchronization and precision both analytically and numerically.
In Section~\ref{sec:many.oscillators}, we consider the case of many oscillators with homogeneous frequencies and present numerical results on their synchronization behavior and their collective precision. Moreover, we study the onset of synchronization in a system with inhomogeneous frequencies.
Finally, in Section~\ref{sec:discussion}, we briefly summarize our results, discuss their relevance, and suggest directions for further studies.

\section{A single phase-discretized oscillator}
\label{sec:single.osc}

\begin{figure}[t]
\begin{center}
\includegraphics[width=8.6cm]{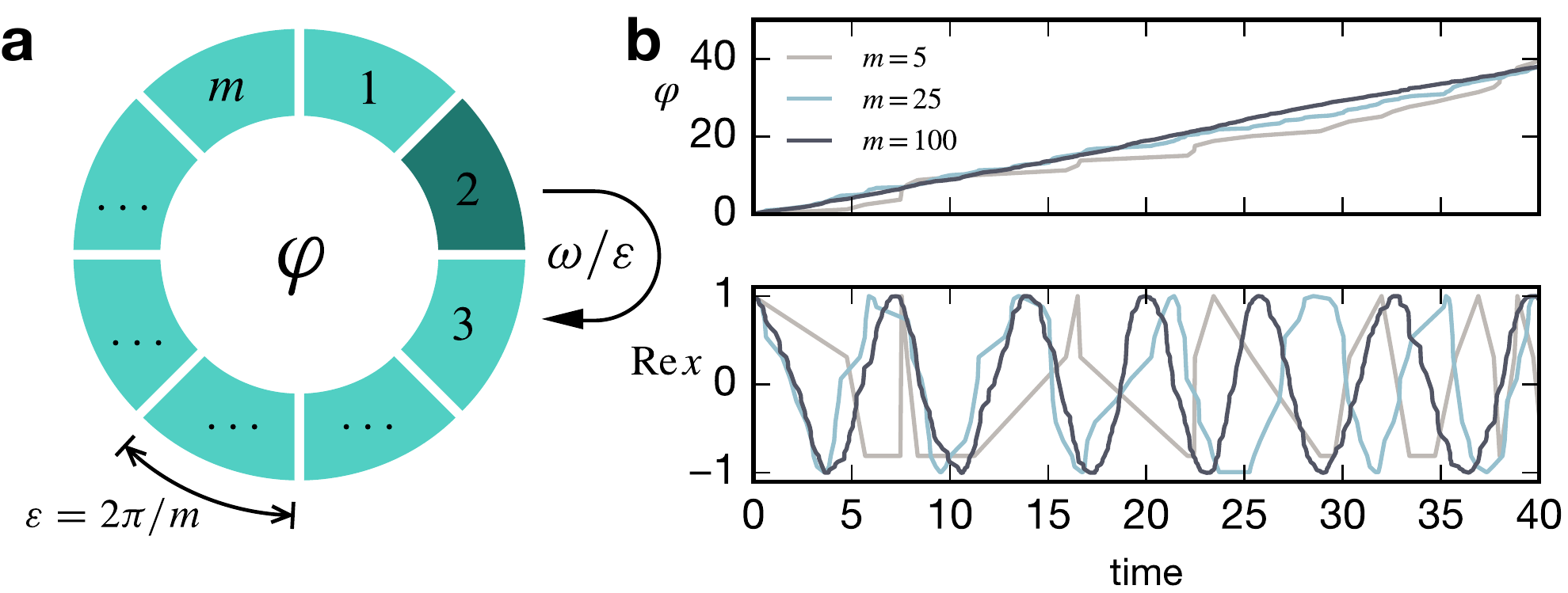}
\caption{(a) Schematic depiction of a phase-discretized oscillator with phase increment $\varepsilon=2\pi/m$ with $m\in\mathds{N}$ and transition frequency~$\omega$. (b) Stochastic trajectories of the phase $\phi=\varepsilon\varphi$ and the oscillatory signal $\operatorname{Re} x = \cos \phi$ for a single oscillator for different values of $m$.}
\label{fig.singleosc}
\end{center}
\end{figure}

\noindent%
We start by considering a single phase-discretized oscillator.
We discretize the phase interval $[0,2\pi)$ into $m$ states and allow the oscillator to advance by discrete phase increments of size $\varepsilon=2\pi/m$, so that its state is given by the discrete phase variable $\varphi \in \mathds{Z}$ (Fig.~\ref{fig.singleosc}a).
The discrete state $\varphi$ is associated with a phase $\phi=\varepsilon\varphi \in \mathds{R}$ and the corresponding oscillatory signal $x=\exp(\mathrm{i}\phi)$.

The stochastic dynamics of the oscillator is governed by a master equation for the probability $P=P(\varphi,t)$ that the oscillator has the discrete phase $\varphi$ at time $t$.
Introducing a transition frequency $\omega \geq 0$, we describe the transition $\varphi \to \varphi+1$ as a Poisson process with transition rate $\omega/\varepsilon$ for a given discretization $m$ (Fig.~\ref{fig.singleosc}a). This ensures that the average duration of one revolution is given by $2\pi/\omega$.
The corresponding master equation is given by
\begin{align}
	\varepsilon \frac{\partial P}{\partial t} &= \omega P(\varphi - 1,t)- \omega P(\varphi,t) \ . \label{eq.single.osc}
\end{align}
The solution to Eq.~(\ref{eq.single.osc}) for the initial condition $P(\varphi,0)=\delta_{\varphi\varphi'}$ is a Poisson distribution \cite{vanKampen2011}, 
\begin{align}
	P(\varphi,t|\varphi',0) = \mathrm{Poisson}(\omega t / \varepsilon,\varphi-\varphi') \ , \label{eq.Poisson}
\end{align}
where $\mathrm{Poisson}(\lambda, n)=\lambda^n \smash{\mathrm{e}^{-\lambda}} \Theta(n)/n!$ with $\Theta$ being the Heaviside function.
Fig.~\ref{fig.singleosc}b shows examples of stochastic trajectories for different phase discretizations~$m$, obtained from a standard stochastic simulation algorithm \cite{Gillespie1977}.

 To determine the dynamical frequency of the oscillator and the frequency fluctuations introduced by stochasticity, we compute the temporal autocorrelation function $G(t) = \langle x(t) x^*(0) \rangle$ of the associated oscillatory signal, where the star denotes the complex conjugate. For the system specified by Eq.~(\ref{eq.single.osc}), it assumes the form $G(t) = \exp(\mathrm{i}\tilde\omega t-k t)$, where the effective frequency $\tilde\omega$ and the decorrelation rate $k$ are given by
\begin{align}
	\tilde\omega = \frac{\sin \varepsilon}{\varepsilon} \omega \ , \quad\quad k = \frac{1-\cos\varepsilon}{\varepsilon} \omega \ .
	\label{eq.eff}
\end{align}
Note that both the effective frequency and the decorrelation rate are proportional to $\omega$, which is the only (inverse) time scale in the system.
Notably, the effective frequency $\tilde\omega$ is systematically smaller than the transition frequency $\omega$.
This difference is due to a `stroboscopic' effect: 
Starting from a defined state $\varphi'$ at time 0, Eq.~(\ref{eq.Poisson}) implies that the discrete phase increment $\Delta\varphi = \varphi-\varphi'$ at $t>0$ is Poisson-distributed with mean and variance depending on $m$ through $\varepsilon$.
For coarse phase discretizations $m$, the tail of the Poisson distribution can considerably extend into regions with phase increments $\Delta\varphi>m$; that is, for a given elapsed time interval, there is a non-vanishing probability for the oscillator to advance by more than one complete cycle.
However, since the oscillatory signal $x$ is $m$-periodic in $\varphi$, a phase increment $\Delta\varphi$ larger than $m$ leads to the same signal as the increment $(\Delta\varphi \operatorname{mod} m)<m$, implying that in this case, the oscillator seemingly goes more slowly.
As expected, in the limit of a continuous phase $\varepsilon \to 0$, we recover $\tilde\omega \to \omega$ and $k \to 0$.

Frequency fluctuations are commonly characterized by the quality factor $Q = \tilde\omega/(2\pi k)$ \cite{Pikovsky1997,Morelli2007}, which is independent of the absolute frequency scale and corresponds to the number of oscillations over which the oscillatory signal stays correlated,
\begin{align}
	Q = \frac{1}{2\pi \tan (\varepsilon/2) } \ . \label{eq.quality.1}
\end{align}
For large phase discretizations $m$, the quality factor quickly approaches the asymptotic behavior
\begin{align}
	Q = \frac{m}{2\pi^2} - \frac{1}{6m} + \mathcal{O}(m^{-3}) \label{eq.quality.expansion}
\end{align}
and becomes effectively linear in $m$. Hence, even to achieve a very low quality factor of $Q=1$, about $m=20$ internal states are required.

To make the connection to the Kuramoto model Eq.~(\ref{eq.kuramoto}), we derive the Langevin equation describing the system in the large-$m$ limit by a system size expansion \cite{vanKampen2011}, see Appendix~\ref{app:linear.noise}. 
This yields
\begin{align}
	\frac{\mathrm{d}\phi}{\mathrm{d}t} = \omega + \sqrt{\frac{2\pi\omega}{m}} \eta(t) \ , \label{eq.langevin}
\end{align}
where $\eta$ is Gaussian white noise with $\langle \eta(t) \rangle = 0$ and $\langle \eta(t) \eta(t') \rangle = \delta(t-t')$. The noise strength grows with the transition frequency $\omega$, a reflection of the fact that the quality factor---which in the case of the Langevin equation (\ref{eq.langevin}) is given by the linear term in Eq.~(\ref{eq.quality.expansion})---only depends on the phase discretization $m$ but not the transition frequency $\omega$, which provides the only time scale in the system and, therefore, cannot affect any dimensionless physical quantity.

\section{Description of oscillator coupling}
\label{sec:osc.coupling}

\noindent%
While the formalization of an uncoupled phase-discretized oscillator seems straightforward, introducing coupling to such a system opens a plethora of different possibilities, even if coupling processes are constrained to a set of Poisson processes running in parallel.
A general formulation of oscillator coupling inevitably introduces backward jumps of the phase since the coupling strength may exceed the intrinsic frequency of an oscillator and may therefore lead to a negative dynamic frequency.
While different stochastic formulations can produce the same mean-field dynamics and/or the same phase-continuous limit $m\to\infty$, fluctuations depend on the details of the stochastic dynamics, e.g., whether for each oscillator, (i) forward/backward jumps of the discrete phase are independent processes running in parallel or (ii) whether only forward or only backward processes can occur depending on the phase relation to other oscillators. We give a brief discussion of different possibilities in Appendix~\ref{app:alternative.couplings}.

\subsection{Stochastic Kuramoto model with discretized phases}

\noindent%
With these different possibilities in mind, we can now write a specific stochastic formulation of coupled phase-discretized oscillators.
The probability $P=P(\varphi_1,\hdots,\varphi_N,t)$ of $N$ oscillators with discrete states $\varphi_1,\hdots,\varphi_N \in \mathds{Z}$ is governed by the master equation
\begin{align}
	\varepsilon \frac{\partial P}{\partial t} &= \sum_{i} \bigg\{ \hat\omega_i + \kappa \sum_{j} c_{ij} \hat\Gamma_{ij}(\varphi_j-\varphi_i) \bigg\} P \ , \Bigg.
	\label{eq.me.main}
\end{align}
where the operators $\hat\omega_i$ and $\hat\Gamma_{ij}$ describe the stochastic dynamics of intrinsic oscillations and coupling, respectively.
As in the Kuramoto model Eq.~(\ref{eq.kuramoto}), $\kappa \geq 0$ denotes the coupling strength and $c_{ij}$ is the adjacency matrix determining the coupling topology.
The intrinsic frequency operator $\hat\omega_i$ assumes the generic form
\begin{align}
	\hat\omega_i &=|\omega_i | (\hat\varphi_i^{-{\operatorname{sign}\omega_i}} - 1 ) \ ,
\end{align}
where $\omega_i$ is the transition frequency of oscillator $i$ and where we have used the ladder operator notation $\smash{\hat\varphi_i^\pm}$, defined by \cite{vanKampen2011}
\begin{align}
\smash{\hat\varphi_i^\pm} P(\varphi_1,\hdots,\varphi_N,t)=P(\varphi_1,\hdots,\varphi_i\pm 1,\hdots,\varphi_N,t) \ . \label{eq.ladder}
\end{align}
For a given $2\pi$-periodic coupling function $\Gamma(\phi)$ taking values between $-1$ and $1$, we define the coupling operator as
\begin{align}
\begin{split}
	\hat\Gamma_{ij}(\varphi) &=  \frac{1+\Gamma(\varepsilon[\varphi + 1]) }{2} \hat{\varphi}_i^{-} - 1 \\
	&\qquad + \frac{1-\Gamma(\varepsilon[\varphi - 1])}{2} \hat{\varphi}_i^{+}  \ .
\end{split} \label{eq.cpl.operator}
\end{align}
This specific formulation of the coupling term corresponds to a biased discrete diffusion process on the discretized phase space, where the bias dynamically depends on the phase difference through the coupling function~$\Gamma(\phi)$; i.e., depending on the phase difference, one of the processes $\varphi_i \to \varphi_i+1$ or $\varphi_i \to \varphi_i-1$ is favored.
Note that the coupling operator given by Eq.~(\ref{eq.cpl.operator}) does not explicitly depend on the index $j$ of the sending oscillator as compared to other discretization schemes, see Appendix~\ref{app:alternative.couplings}.

Note that in Eq.~(\ref{eq.me.main}), the expression in parentheses formally resembles the r.h.s.~of the Kuramoto model Eq.~(\ref{eq.kuramoto}) with parameters and functions promoted to Liouville operators.
In general, the phase discretization leads to two major differences to the classical Kuramoto model Eq.~(\ref{eq.kuramoto}): (i) oscillator dynamics is now inherently stochastic and (ii) the coupling function is sampled at discrete readout points determined by the phase discretization.

\subsection{Linear noise approximation}
\label{subsec:lna}

\noindent%
To establish a connection to the classical Kuramoto model Eq.~(\ref{eq.kuramoto}), we carry out a system size expansion of the system described by Eqs.~(\ref{eq.me.main}--\ref{eq.cpl.operator}), formally interpreting the phase discretization $m$ as the system size.
This enables to write a linear noise approximation (LNA) for the corresponding stochastic governing equations for the physical phases $\phi_i = \varepsilon \varphi_i$. Details on the derivation are given in Appendix~\ref{app:linear.noise}.
The resulting linear noise approximation for the physical phase is given by
\begin{align}
	\phi_i(t) = \Phi_i(t) + \sqrt{\frac{2\pi}{m}} \xi_i(t) + \mathcal{O}(m^{-1}) \ , \bigg.  \label{eq.lna.phase} 
\end{align}
where $\Phi_i$ is a `macroscopic' phase variable obeying the deterministic Kuramoto dynamics
\begin{align}
	\frac{\mathrm{d}\Phi_i}{\mathrm{d}t} &= \omega_i  + \kappa \sum_{j} c_{ij}   \Gamma(\Phi_j-\Phi_i) \ , \Bigg. \label{eq.lna.det}
\end{align}
cf.~Eq.~(\ref{eq.kuramoto}), and $\xi_i$ is a random variable which is governed by the Langevin equation
\begin{align}
	\frac{\mathrm{d}\xi_i}{\mathrm{d}t} &= \kappa \sum_{j} c_{ij} \Gamma'(\Phi_j-\Phi_i) (\xi_j-\xi_i) + \sqrt{\mu_i} \eta_i(t) \ , \Bigg.  \label{eq.lna.noise} 
\end{align}
where $\Gamma'$ is the derivative of the coupling function, $\eta_i$ is Gaussian white noise with $\langle \eta_i(t) \rangle = 0$ and $\langle \eta_i(t) \eta_j(t') \rangle = \delta_{ij}\delta(t-t')$, and $\mu_i$ is the effective noise strength for oscillator $i$, given by
\begin{align}
	\mu_i = |\omega_i |+ \kappa c_i \ ,\label{eq.phase.diffusion}
\end{align}
where $c_i=\sum_j c_{ij}$ is the total coupling weight of the oscillators coupled into oscillator $i$.
Eq.~(\ref{eq.phase.diffusion}) illustrates that in our formulation of the phase-discretized system, noise has two sources: the intrinsic oscillatory dynamics, as indicated by the intrinsic transition frequency $\omega_i$ and already shown in Eq.~(\ref{eq.langevin}), but also coupling which, as an inherently stochastic process, inevitably contributes noise to the system.
For a given oscillator~$i$, coupling to each connected oscillator $j$ is an independent process; therefore, the total contribution from coupling to its noise strength is proportional to the total coupling weight $c_i$.
Hence, the net effect of coupling on the synchronization and precision properties of the coupled system are not immediately obvious.
Note that in the limit of a continuous phase $m\to\infty$, the phases behave as $\phi_i \to \Phi_i$ and the classical Kuramoto model Eq.~(\ref{eq.kuramoto}) for the physical phases $\phi_i$ is recovered.

Note also that for the master equation of the single oscillator, Eq.~({\ref{eq.single.osc}}), which has state-independent transition rates, the derived linear noise approximation is exact up to second order in the moments {\cite{Grima2015}}.
On the other hand, it is not a priori obvious under which circumstances the linear noise approximation Eq.~({\ref{eq.lna.phase}}--{\ref{eq.phase.diffusion}}) is a good approximation for the full model Eqs.~({\ref{eq.me.main}}--{\ref{eq.cpl.operator}}), as it involves (i) nonlinearly state-dependent transition rates and (ii) entails an expansion of the coupling function $\Gamma$, suggesting that its validity is constrained to the vicinity of states for which $\Gamma$ is approximately linear around the occurring phase differences. In Section~{\ref{sec:many.oscillators}}, we demonstrate its effectiveness in describing steady-state properties by numerical simulations.

\section{Dynamics of two coupled oscillators}
\label{sec:two.oscillators}

\noindent%
To gain some insights into the stochastic behavior of the coupled system, we first investigate the simplest case of two coupled oscillators without self-coupling, $N=2$ and $c_{ij}=1-\delta_{ij}$.
For concreteness, we consider the generic class of coupling functions of the Kuramoto--Sakaguchi type in the following \cite{Sakaguchi1988}, 
\begin{align}
	\Gamma(\phi) = \sin (\phi-\phi_0) \ , \label{eq.KS.coupling}
\end{align}
where $\phi_0 \in [0,2\pi)$ is a constant phase shift.
First, we address the transient behavior of the oscillators approaching synchrony, followed by an analysis of the synchronization and precision properties of the steady state.

\subsection{Synchronization transient}

\begin{figure}[t]
\begin{center}
\includegraphics[width=8.6cm]{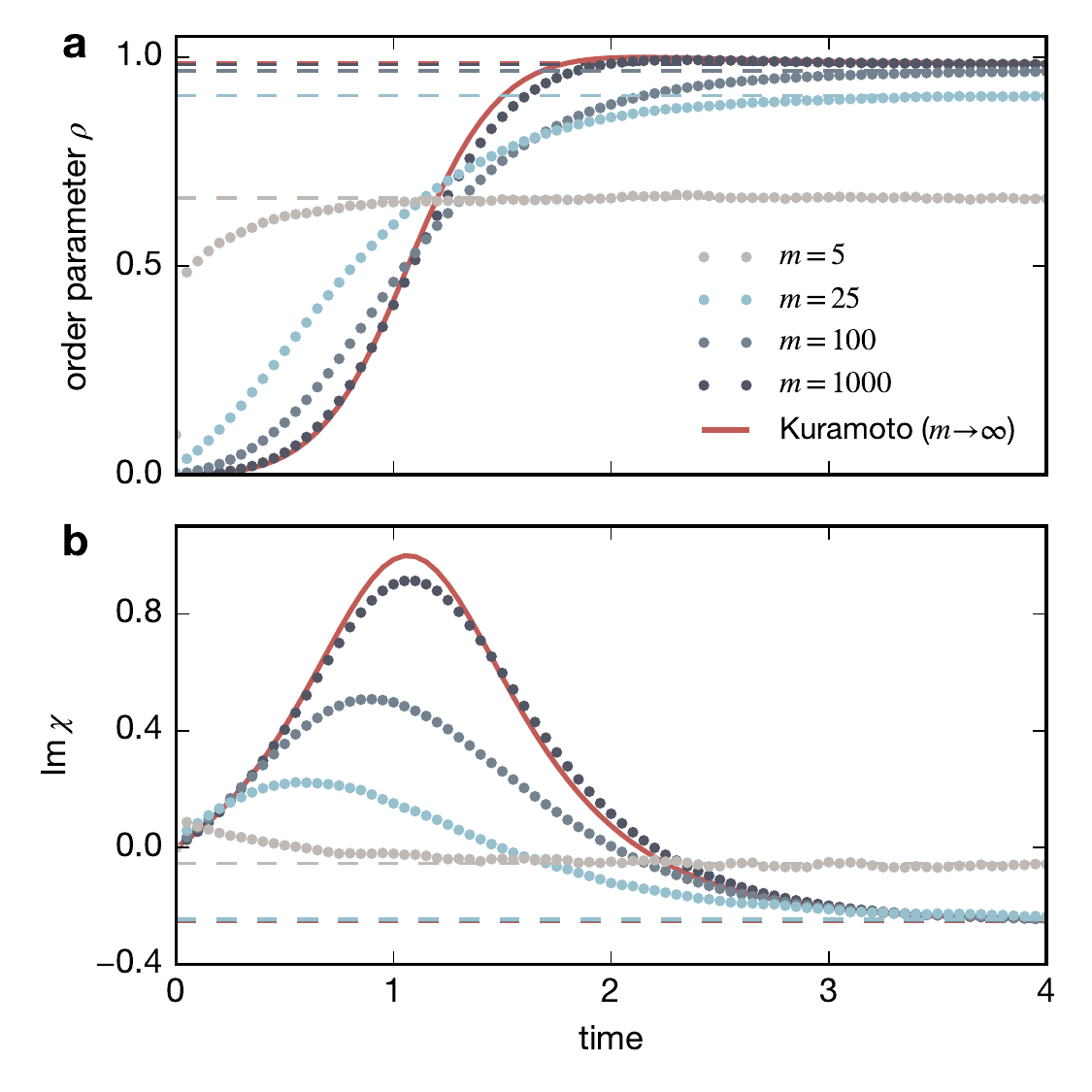}
\caption{Synchronization transient for different values of $m$  for two coupled oscillators. (a) Kuramoto order parameter $\rho$, defined in Eq.~(\ref{eq.order}); (b) imaginary part of the cross correlation $\chi(t) = \langle x_1(t) x_2^*(t) \rangle$. Dots show averages over stochastic trajectories of the phase-discretized model Eq.~(\ref{eq.me.main}) with the coupling function Eq.~(\ref{eq.KS.coupling}), the red solid line shows the result for the classical Kuramoto model Eq.~(\ref{eq.kuramoto}), dashed horizontal lines show the exact steady-state solutions, given by Eqs.~(\ref{eq.ss}) and (\ref{eq.cc.ss}), respectively. Parameters are $\omega_1 = 0.75$, $\omega_2 = 1.25$, $\kappa=1$, $\phi_0=0$. 
}
\label{fig.sync}
\end{center}
\end{figure}

\noindent 
The time-dependent phase coherence of the two oscillators can be monitored via the Kuramoto order parameter, defined by \cite{KuramotoBook1984}
\begin{align}
	\Psi(t) = \frac{1}{N}  \sum_{i=1}^N x_i(t) \ , \label{eq.kuramoto.op}
\end{align}
where $N$ is the number of oscillators and $x_i(t)=\mathrm{e}^{\mathrm{i}\varepsilon\varphi_i(t)}$ is the oscillatory signal associated with oscillator $i$, as before. Usually, one considers the magnitude $|\Psi|$, which takes values from $0$ to $1$ with $|\Psi|=1$ indicating perfect phase coherence and $|\Psi|<1$ indicating the existence of phase lags between oscillators. Here we focus on the squared magnitude $|\Psi|^2$, which basically has the same interpretation but simpler analytical properties. For two oscillators, the cross correlation $\chi(t) = \langle x_1(t) x_2^*(t) \rangle$
contains the expectation value of $|\Psi|^2$ in its real part,
\begin{align}
	\rho(t) \equiv \langle |\Psi(t)|^2 \rangle = \frac{1+\operatorname{Re} \chi(t)}{2} \ . \label{eq.order}
\end{align}
Since $|\Psi|^2$ is bounded, a value of $\rho$ close to 1 indicates not only a small average phase difference but also small fluctuations in the phase difference.
Fig.~\ref{fig.sync} shows both $\rho(t)$ as well as $\operatorname{Im}\chi(t)$ (which together carry the same information as the full cross correlation $\chi$)
for different phase discretizations~$m$ for two oscillators with unequal frequencies and an initially maximally desynchronized state.
After an initial transient, the system approaches a steady state with a constant order parameter and cross correlation
\begin{align}
	R &= \lim_{t\to\infty} \rho(t) \ , \qquad {X} = \lim_{t\to\infty} \chi(t) \ , \label{eq.ss.ind}
\end{align}
which depend on $m$.

\begin{figure}[t]
\begin{center}
\includegraphics[width=8.6cm]{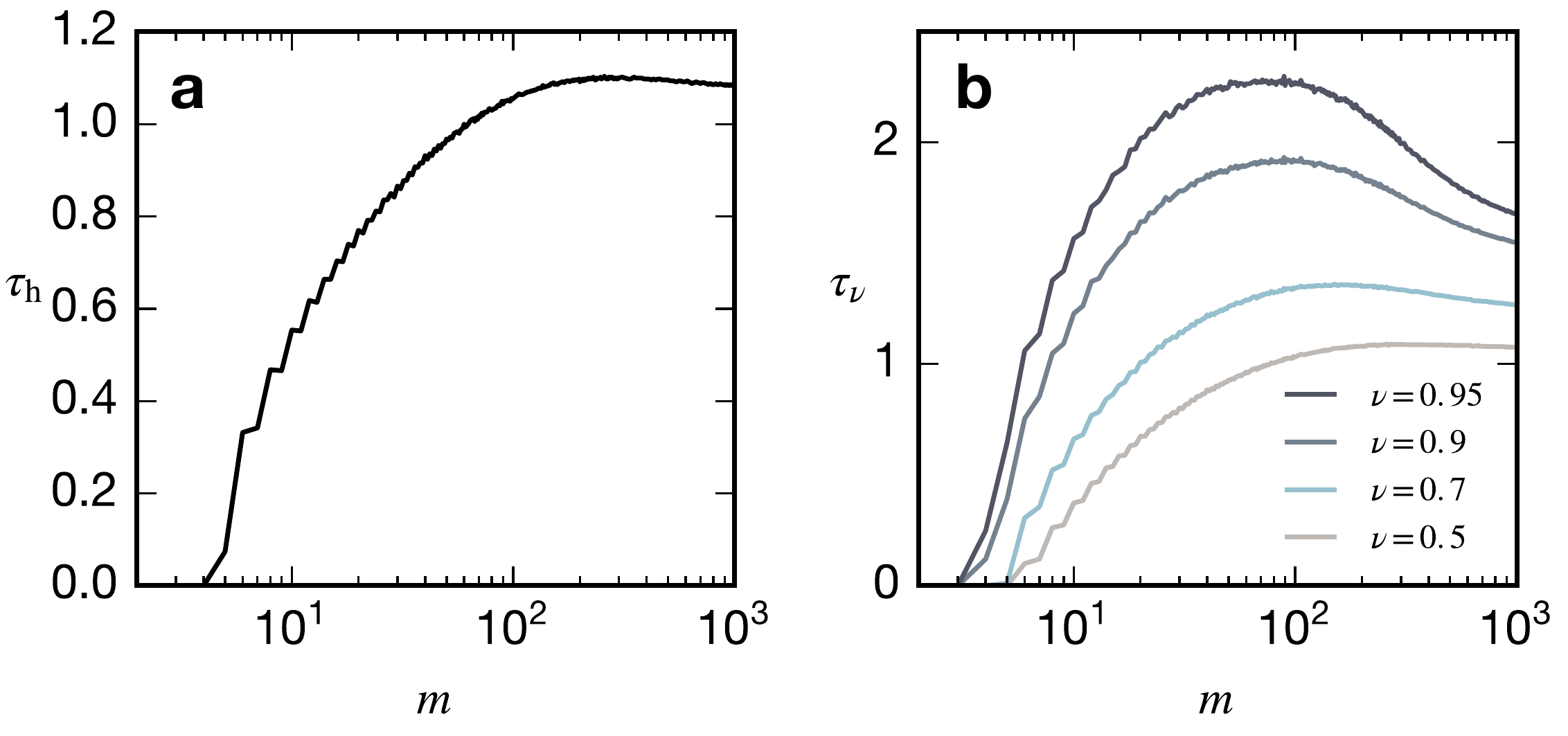}
\caption{Synchronization time towards the steady state as a function of the phase discretization $m$  for two coupled oscillators. (a) Time $\tau_\mathrm{h}$ it takes to reach the order parameter $1/2$, defined by Eq.~(\ref{eq.sync.time.abs}); (b) time $\tau_\nu$ it takes to reach a fraction $\nu$ of the steady-state order parameter $R$, defined by Eq.~(\ref{eq.sync.time.rel}).
System parameters as in Fig.~\ref{fig.sync}.
}
\label{fig.sync.time}
\end{center}
\end{figure}

\begin{figure*}[t]
\begin{center}
\includegraphics[width=18cm]{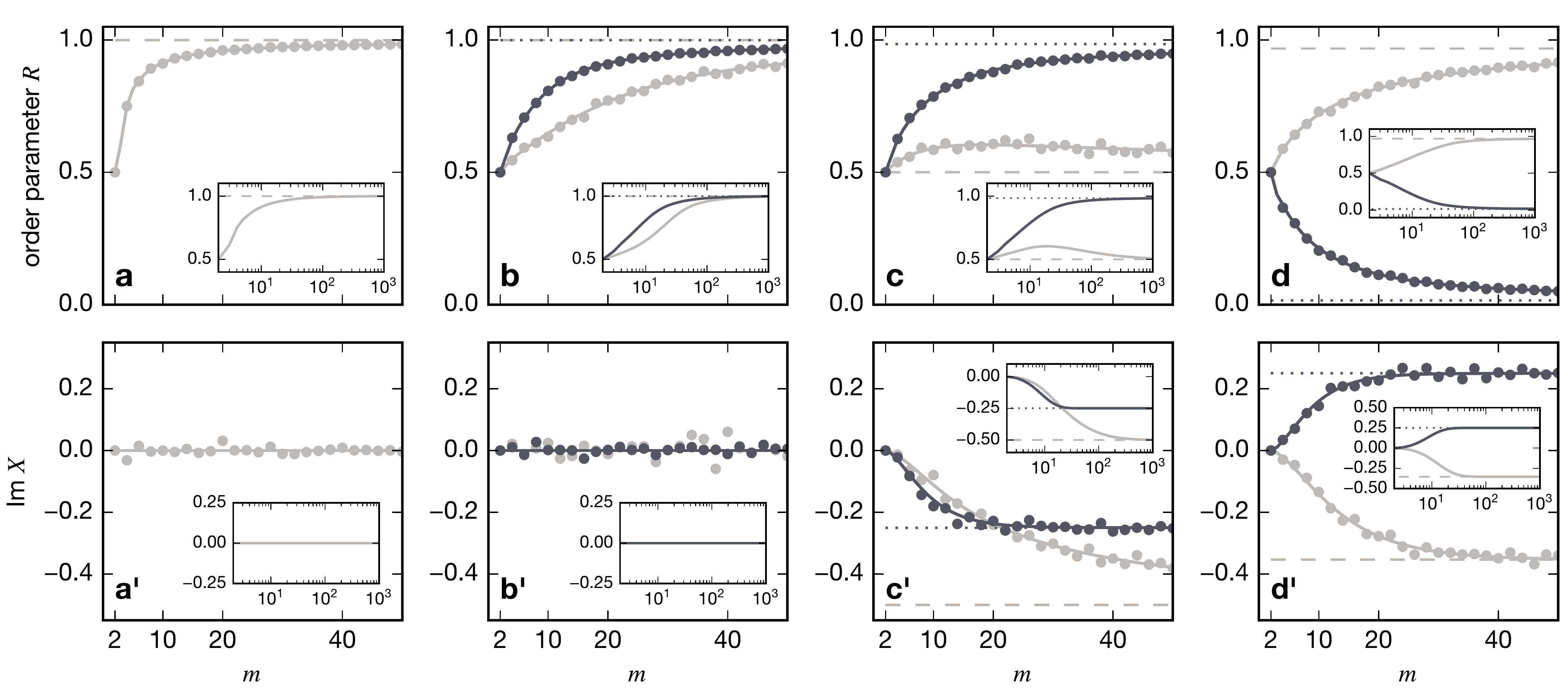}
\caption{Phase coherence of the two-oscillator system for different phase discretizations. Plots show the expectations values of the steady-state order parameter $R$ and the imaginary part of the cross correlation ${X}$ for different configurations. Solid lines show the exact solutions, given by Eqs.~(\ref{eq.ss}) and (\ref{eq.cc.ss}), respectively; dots show time averages of simulated stochastic trajectories of length $T=2000$; dashed and dotted horizontal lines show the corresponding results for the classical Kuramoto model. (a,a') Zero intrinsic frequencies: $\omega_1=\omega_2=0$, $\kappa=1$, $\phi_0=0$; (b,b') Equal intrinsic frequencies: $\omega_1=\omega_2=1$, $\phi_0=0$, for $\kappa=1$ (dark) and $\kappa=0.25$ (light); (c,c') Unequal frequencies: $\omega_1=0.75$, $\omega_2=1.25$, $\phi_0=0$ with $\kappa=1$ (dark) and $\kappa=0.2$ (light); (d,d') Non-zero phase shift: $\omega_1=0.75$, $\omega_2=1.25$, $\kappa=1$, for $\phi_0=\pi$ (dark) and $\phi_0=\pi/4$ (light). Insets show the corresponding exact solutions for a larger range of the phase discretization $m$ (note the logarithmic scale of the $m$-axis).}
\label{fig.steady.state}
\end{center}
\end{figure*}

Even though coarser phase discretizations typically entail a lower degree of synchrony at steady-state, such systems tend to initially synchronize faster than system with a finer discretization (see Fig.~\ref{fig.sync}a). To illustrate this behavior, we define, for a system starting from the completely desynchronized state with maximum phase difference, as two complementary quantities the time $\tau_\mathrm{h}$ it takes for the order parameter to reach the absolute value $1/2$ and the time $\tau_\nu$ it takes to reach a relative fraction $\nu$ of the steady-state order parameter~$R$,
\begin{align}
	\tau_\mathrm{h} &= \min \{ t \ | \ \rho(t) \geq 1/2 \}  \ , \label{eq.sync.time.abs} \\
	\tau_\nu &= \min \{ t \ | \ \rho(t) \geq \nu R \}  \ . \label{eq.sync.time.rel}
\end{align}
Fig.~\ref{fig.sync.time} shows the synchronization times as a function of the phase discretization $m$ for different values of $\nu$ and reveals an interesting behavior:
The time to reach the absolute order parameter $\rho=1/2$ tends to decrease for coarser phase discretizations even though the transition frequencies $\omega_i$ and the coupling strength $\kappa$ are kept constant (Fig.~\ref{fig.sync.time}a).
In contrast, the time to reach a relative fraction of the steady-state order parameter attains a distinct maximum for finite discretizations (Fig.~\ref{fig.sync.time}b).
Therefore, coarser phase discretizations can facilitate faster initial synchronization even though they eventually reach a smaller phase-coherence and take a longer time reach the vicinity of their steady state.

\subsection{Steady-state phase coherence}

\noindent
How does the steady-state phase coherence depend on the phase discretization? And how does this compare to the synchronized state of two coupled Kuramoto oscillators with detuning?
Let us briefly recapitulate some results from the classical Kuramoto model \cite{Adler1946,Schuster1989}. There, the system assumes a phase-locked steady state if coupling is strong enough to overcome the frequency difference between the oscillators, that is, if $|\gamma|<1$ where
\begin{align}
	\gamma=\frac{|\omega_1-\omega_2|}{2\kappa \cos\phi_0} \ .\label{eq.freq.diff}
\end{align}
In this case, the order parameter is given by
\begin{align}
	R = \frac{1+\operatorname{sign}(\gamma)  \sqrt{1- \gamma^2}}{2} \ .\label{eq.order.Kuramoto}
\end{align}
Hence, in terms of the intrinsic frequences, the order parameter is determined by the absolute frequency difference $|\omega_1-\omega_2|$ in a monotonic way.
For $|\gamma|> 1$, both oscillators phase-drift with respect to each other and the time average of the order parameter is $1/2$.

In the case of the phase-discretized model, nonlinear coupling combines with stochasticity and therefore, an analysis is more involved.
Nevertheless, an exact solution for the steady-state order parameter $R$ and the cross correlation $X$, Eqs.~(\ref{eq.ss.ind}), can be constructed, see Appendix~\ref{appendix:exact.signal} for a derivation.
Without loss of generality, we consider the case $\omega_1 \geq 0$, $\omega_2\geq 0$ for which the resulting order parameter is given by
\begin{align}	
	R &= \frac{1}{2} \bigg( 1 + \operatorname{Re} \bigg\{ \prod_{n=1}^{m-1}\Lambda_n -  \Lambda_1 \bigg\} \bigg)  \ , \label{eq.ss}
\end{align}
where $\Lambda_n$ can be represented as the continued fraction
\begin{align}
	\Lambda_{n} = -\operatornamewithlimits{\mathbf{K}}_{i=n}^{m-1} \lambda_i \equiv -\frac{1}{\lambda_{n}+ \frac{1}{\lambda_{n+1} + \frac{1}{\ddots + \frac{1}{\lambda_{m-1}}}}}\label{eq.contfrac}
\end{align}
with
\begin{align}
\lambda_n
	&= \frac{(\omega_1+\omega_2+2\kappa ) \tan (\pi n/m ) - \mathrm{i}(\omega_1-\omega_2)}{\kappa \cos\phi_0} \ .
	\label{eq.eta}
\end{align}
Interestingly, the order parameter $R$ depends not only on the frequency difference $\omega_1-\omega_2$ but also on the frequency sum $\omega_1+\omega_2$ through $\lambda_n$. This reflects the fact that in the stochastic system, the degree of noise depends on the frequency scale (cf.~Eq.~(\ref{eq.langevin}) and the discussion below).
Due to its combinatorial complexity, the exact solution given by Eq.~(\ref{eq.ss}--\ref{eq.eta}) is somewhat opaque; therefore, we give a few explicit expressions for small phase discretizations~$m$ in Appendix~\ref{appendix:exact.signal}.

\begin{figure}[t]
\begin{center}
\includegraphics[width=8.3cm]{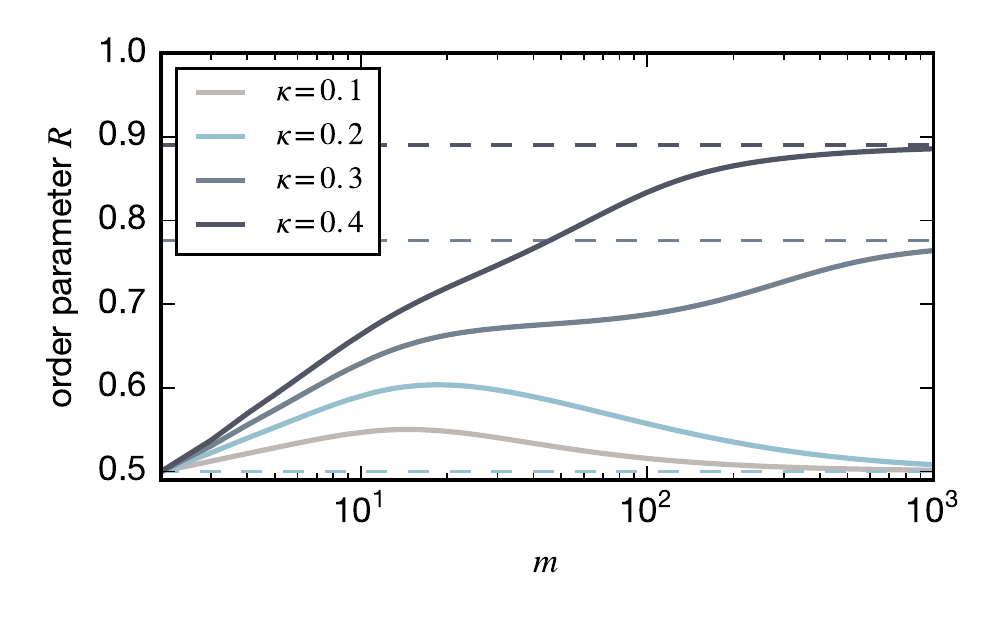}
\vspace{-0.7cm}
\caption{Steady-state order parameter $R$ as a function of the phase discretization $m$ for different coupling strengths $\kappa$  for two coupled oscillators, as given by Eq.~(\ref{eq.ss}). Dashed lines show the Kuramoto limit $m\to\infty$, given by Eq.~(\ref{eq.order.Kuramoto}). The other parameters are $\omega_1=0.75$, $\omega_2=1.25$, $\phi_0=0$ as in Fig.~\ref{fig.steady.state}c.}
\label{fig.nonmonotonic}
\end{center}
\end{figure}

Fig.~\ref{fig.steady.state} shows the order parameter $R$ and the imaginary part of the cross correlation ${X}$ as a function of the phase discretization $m$ for different frequency detunings and coupling strengths, both from numerical simulations of stochastic trajectories (dots) and the exact solutions given by Eqs.~(\ref{eq.ss}) and (\ref{eq.cc.ss}) (solid lines).
For many generic parameter combinations, the order parameter monotonically increases with finer phase discretizations.
However, in a few cases, the behavior of the order parameter and the cross correlation show some remarkable features.
First, even at coupling strengths below the classical critical value that ensures $|\gamma|< 1$ we detect partial synchrony, i.e., an order parameter $R > 1/2$ (bright curve in Fig.~{\ref{fig.steady.state}}c, corresponding to $\gamma=1.25$), indicating that the system spends a larger time in regions with small phase differences.
Second, while in all cases the order parameter approaches the Kuramoto value in the limit $m\to\infty$, the convergence is not always monotonic. In fact, there are phase discretizations $m$ for which the degree of partial synchrony becomes maximal. This is exemplified by the bright curve in Fig.~{\ref{fig.steady.state}}c and in Fig.~\ref{fig.nonmonotonic}, where the order parameter is displayed for different coupling strengths and up to very fine phase discretizations. The curves below the critical coupling strength $\kappa_\mathrm{c} = |\omega_1-\omega_2|/2$ exhibit a non-monotonic behavior with a distinct maximum for a finite phase discretization.

This behavior can be illuminated as follows: In the deterministic case $m\to\infty$, the phase difference $\psi=\phi_1-\phi_2$ of both oscillators is governed by the Adler equation $\mathrm{d}\psi/\mathrm{d}t = -\mathrm{d}v/\mathrm{d}\psi$ with $v(\psi)= -(\omega_1-\omega_2)\psi - 2\kappa \cos\psi$ \cite{Adler1946}, where for simplicity, we have considered the case of zero coupling phase shift, $\phi_0=0$.
Therefore, the phase difference $\psi$ can be interpreted as the position of an overdamped particle moving in the tilted washboard potential $v(\psi)$ \cite{Lindner2004,Shlomovitz2014}.
The dynamic drift velocity $-\mathrm{d}v/\mathrm{d}\psi$ is symmetric around phase differences $\psi_n = (2n+1)\pi/2$ with $n\in\mathds{Z}$, which correspond to an order parameter of $\rho=1/2$. Therefore, if averaged over time, contributions from order parameters larger and smaller than $1/2$ exactly cancel out.
In the case of finite phase discretizations $m$, the system is stochastic and it tends to spend a larger time in states with $\rho>1/2$.
The reason for this can be understood by considering the Adler equation in the presence of noise and interpreting it as the governing equation of an overdamped Brownian particle in the potential $v(\psi)$.
(In the case of the phase-discretized system, we may think of a `discrete' potential whose increments determine the transition rates between states with different discrete phase differences.)
For subcritical coupling strengths $\kappa < \kappa_\mathrm{c}$, the potential $v$ is (i)~monotonic in $\psi$, (ii)~convex in regions with $\rho>1/2$, and (iii)~concave in regions with $\rho<1/2$; the latter can be seen by rewriting its second derivative as a function of the order parameter, $\mathrm{d}^2 v/\mathrm{d}\psi^2=4\kappa(\rho-1/2)$.
Therefore, the particle leaves regions with $\rho<1/2$ on the steepest slope of the potential, making it unlikely to return into the regions due to fluctuations whereas it leaves regions with $\rho>1/2$ where the potential is most shallow, rendering return events due to fluctuations much more likely.

\subsection{Oscillator precision at steady state}
\label{subsec:precision.2osc}

\begin{figure}[t]
\begin{center}
\includegraphics[width=8.6cm]{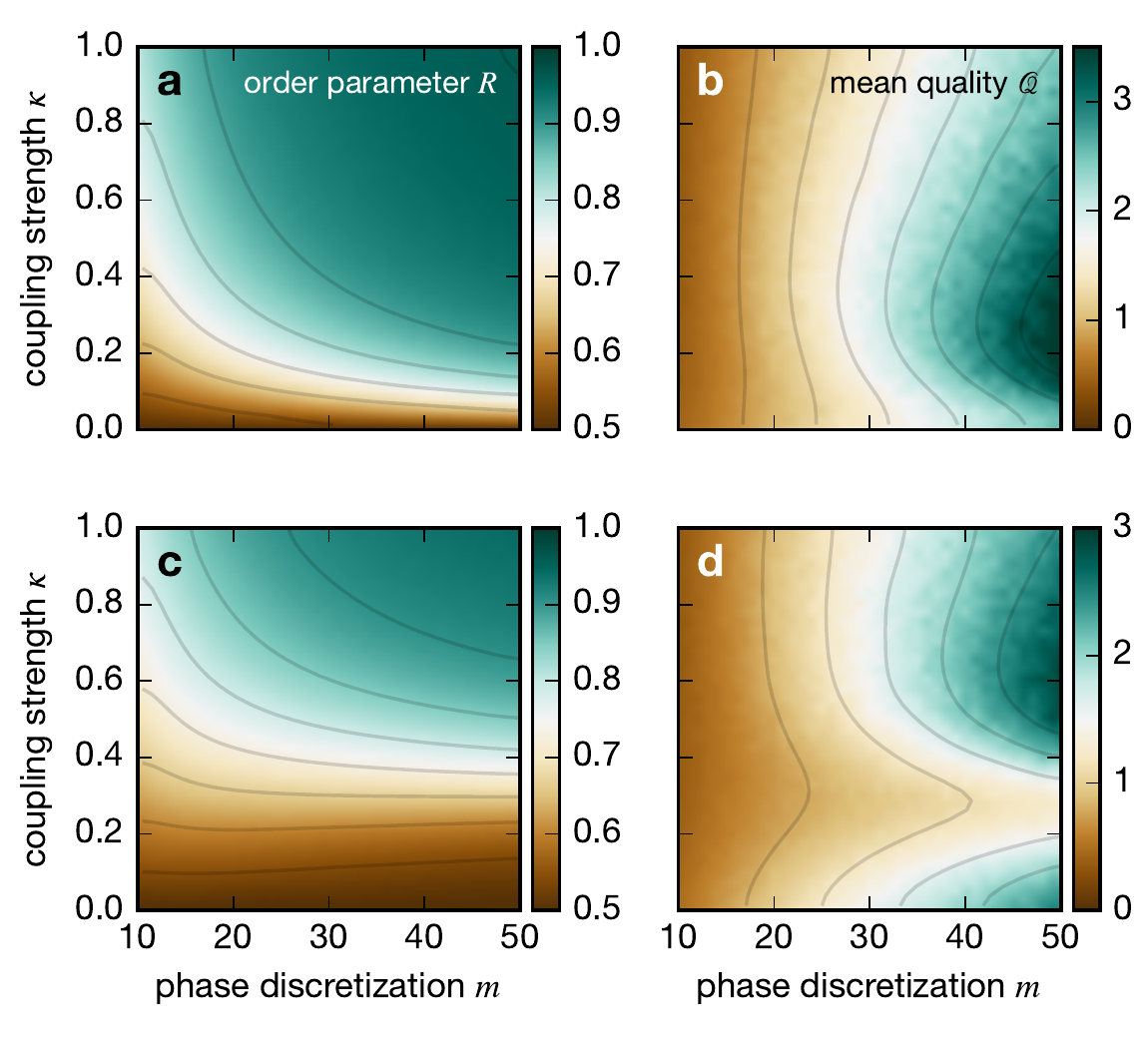}
\caption{Order parameter $R$ (left column) and mean quality $\mathscr{Q}$ (right column) as a function of the phase discretization $m$ and the coupling strength $\kappa$ for two coupled oscillators. The panels show the two cases of (a,b) equal frequencies, $\omega_1=\omega_2=1$; and (c,d) unequal frequencies, $\omega_1=0.75$, $\omega_2=1.25$. The coupling phase shift is $\phi_0=0$.}
\label{fig.quality}
\end{center}
\end{figure}

\noindent%
It is well-known that besides promoting synchronization, coupling can lead to an improvement of the oscillator precision, i.e., often damps frequency fluctuations \cite{Cross2012}.
However, in the phase-discretized system, coupling not only tends to synchronize oscillators but is itself also a source of noise (cf.~Eq.~(\ref{eq.phase.diffusion}) and the discussion below). Hence, the effects of coupling on oscillator precision are not immediately obvious.
To quantitatively assess these effects, we again consider the quality factor of the oscillators (see Section~\ref{sec:single.osc}), now for the coupled case:
from the numerically obtained autocorrelation functions $G_i(t) = \langle x_i(t) x_i^*(0) \rangle$ of the two oscillators $i=1,2$, we obtain the quality factors by a fit with the exponential $\exp(\mathrm{i}\tilde\omega_i t - k_i t)$ as $Q_i=\tilde\omega_i/2\pi k_i$. From this, we compute the mean quality factor $\mathscr{Q} = (Q_1+Q_2)/2$ as a proxy for the quality of the coupled system.
Fig.~\ref{fig.quality} shows the steady-state order parameter $R$ and the steady-state quality factor $\mathscr{Q}$ as a function of the phase discretization~$m$ and the coupling strength~$\kappa$ for the case of equal frequencies (Fig.~\ref{fig.quality}a,b) and the case of unequal frequencies (Fig.~\ref{fig.quality}c,d).
Remarkably, while the order parameter $R$ follows the general trends studied in the previous section, the quality factor $\mathscr{Q}$ exhibits certain optima along the coupling strength axis.
In the case of equal frequencies (Fig.~\ref{fig.quality}a,b), for a given phase discretization, increasing the coupling strength beyond the optimal value contributes more noise to the system than coupling is reducing.
The location of this optimum depends on the intrinsic frequencies of the oscillators and for detuned frequencies, we consequently find two optima along the coupling strength axis (Fig.~\ref{fig.quality}d).
It is also interesting to note that there is no obvious correlation between synchrony and precision along the coupling strength axis, so that a high degree of phase synchrony can indeed be accompanied by large frequency fluctuations.

\section{Synchronization of many oscillators}
\label{sec:many.oscillators}

\noindent%
We now turn to the dynamics of systems with larger numbers of oscillators and choose the classical case of an all-to-all coupled system to illustrate their behavior. 
For a system without self-coupling, the corresponding normalized adjacency matrix is given by $c_{ij}=(N-1)^{-1}(1-\delta_{ij})$.

\subsection{`Mean-field' formulation of the all-to-all coupled system}

\begin{figure*}[t]
\begin{center}
\includegraphics[width=18cm]{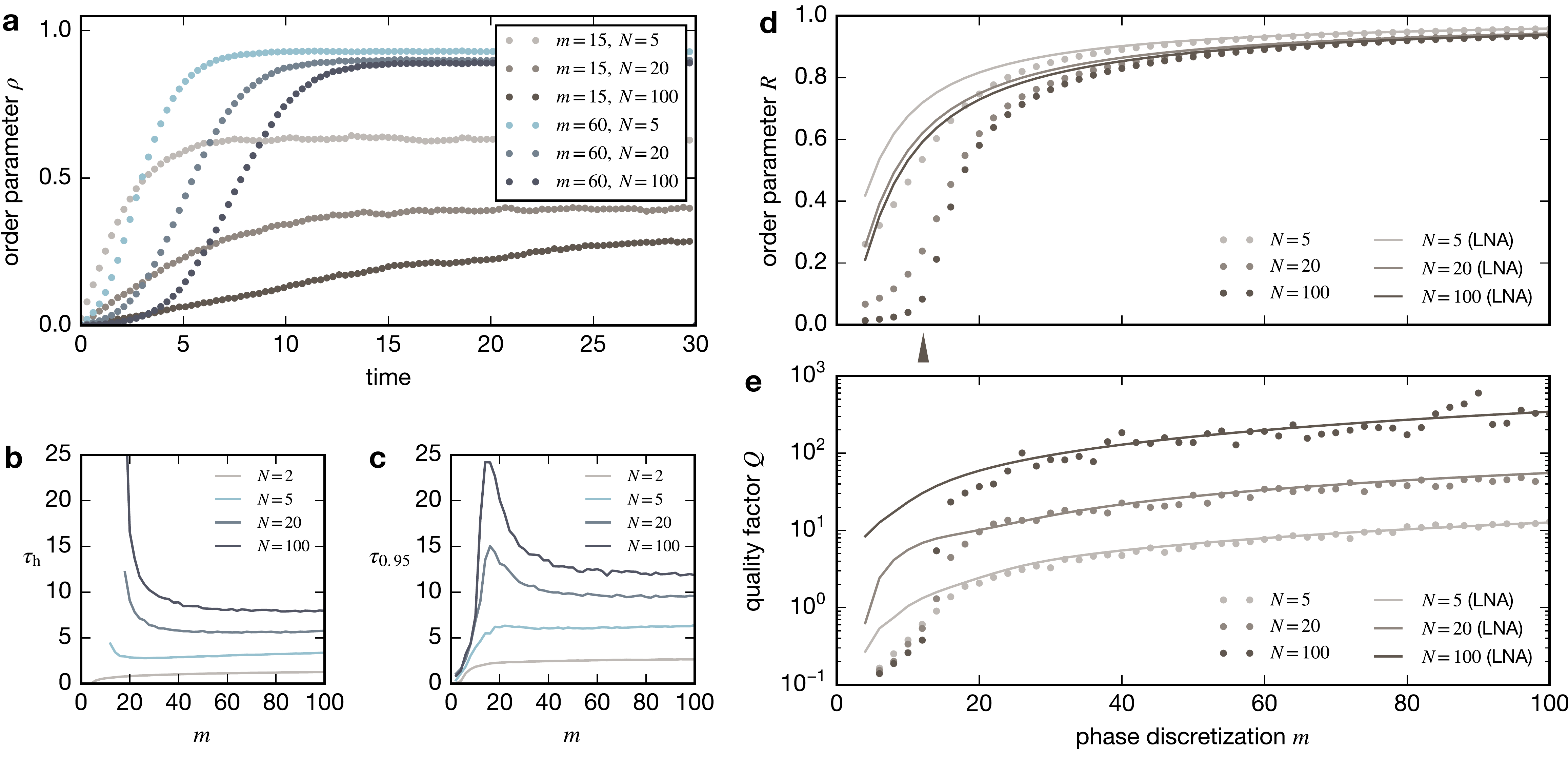}
\caption{Synchronization and precision properties for many-oscillator systems. (a) Synchronization transient as indicated by the time-dependent Kuramoto order parameter $\rho$, Eq.~(\ref{eq.order}), for different phase discretizations and numbers of oscillators. (b,c) Synchronization times as defined in Eqs.~(\ref{eq.sync.time.abs}) and (\ref{eq.sync.time.rel}) as a function of the phase discretization for different numbers of oscillators, analogous to Fig.~\ref{fig.sync.time} for the case of two oscillators. (d,e) Steady-state order parameter $R$ and quality factor $Q$ as a function of the phase discretization for different numbers of oscillators for the full phase-discrete system (dots) and the linear noise approximation (curves), given by Eqs.~(\ref{eq.lna.2}--\ref{eq.lna.ops}). The $m$-axes in panels d and e are the same. System parameters are $\omega_i=1$, $\kappa=1$, $\phi_0=0$.}
\label{fig.many.oscillators}
\end{center}
\end{figure*}

\noindent%
For an all-to-all coupling topology, the original Kuramoto model with sinusoidal coupling function can be rewritten in such a way that each oscillator individually couples to the order parameter $\Psi$, also called the `mean-field' {\cite{Strogatz2000}}.
The same is possible for the phase-discretized stochastic system specified by Eqs.~(\ref{eq.me.main}--\ref{eq.cpl.operator}) and (\ref{eq.KS.coupling}), which can be rewritten in the form\footnote{
The rewriting in the `mean-field' form relies on the fact that for sinusoidal coupling functions, the coupling term factorizes into a term containing the phase of the reference oscillator and the sum over all neighboring oscillators, e.g., $\smash{\sum_j \sin(\phi_i-\phi_j)} = \smash{\operatorname{Im}(\mathrm{e}^{\mathrm{i}\phi_i} \sum_j \mathrm{e}^{-\mathrm{i}\phi_j})} = N \operatorname{Im}(\mathrm{e}^{\mathrm{i}\phi_i} \Psi)$ where $\Psi$ is the order parameter Eq.~(\ref{eq.kuramoto.op}). The same rewriting can be applied to the transition rates of the phase-discretized system specified by Eqs.~(\ref{eq.cpl.operator}) and (\ref{eq.KS.coupling}), which enables the representation Eqs.~(\ref{eq.gc.model}) and (\ref{eq.gc.cpl}).
}
\begin{align}
	\varepsilon \frac{\partial P}{\partial t} &= \sum_{i} \Big\{ \hat\omega_i + \kappa \hat{\Gamma}^\mathrm{MF}_i(\varphi_i,\Psi)  \Big\} P \ , \bigg.
	\label{eq.gc.model}
\end{align}
where $\Psi$ is the Kuramoto order parameter defined in Eq.~(\ref{eq.kuramoto.op}) and the coupling operator $\smash{\hat{\Gamma}^\mathrm{MF}_i}$ is given by
\begin{align}
\begin{split}
	\hat{\Gamma}^\mathrm{MF}_i(\varphi,\Psi) &=\sum_{\sigma=\pm} \left[  \operatorname{Im} \left\{  \frac{N\Psi \mathrm{e}^{-\mathrm{i}\varepsilon\varphi}-1}{N-1} \mathrm{e}^{\mathrm{i}(\sigma\varepsilon-\phi_0)} \right\} \right]_\sigma \hat{\varphi}_i^{-\sigma} \\
	&\qquad - 1 \ ,
\end{split}	\label{eq.gc.cpl}
\end{align}
where we have introduced the notation $[x ]_\pm \equiv (1 \pm x)/2$.
Note that this rewriting also drastically reduces the computational effort to simulate the model\footnote{For stochastic simulations, the advantage of the form given by Eqs.~(\ref{eq.gc.model}) and (\ref{eq.gc.cpl}) is that in order to compute all reaction propensities, it is sufficient to compute the order parameter $\Psi$ and the quantity given by Eq.~(\ref{eq.gc.cpl}) for each of the $N$ phases instead of computing all $N(N-1)$ pairwise phase differences}.

Likewise, the corresponding linear noise approximation Eqs.~(\ref{eq.lna.phase}--\ref{eq.phase.diffusion}) can be recast in the form
\begin{align}
	\frac{\mathrm{d}\Phi_i}{\mathrm{d}t} &= \omega_i  + \frac{\kappa}{1-N^{-1}} r \sin(\psi-\Phi_i-\phi_0)  \ , \label{eq.lna.2} \\[4pt]
\begin{split}
	\frac{\mathrm{d}\xi_i}{\mathrm{d}t} &= \frac{\kappa}{1-N^{-1}} \Big\{ \tilde{r} \cos(\tilde{\psi}-\Phi_i-\phi_0) \\
	&\qquad - r \xi_i \cos(\psi-\Phi_i-\phi_0) \Big\}  + \sqrt{\mu_i} \eta_i(t) \ ,
\end{split} \label{eq.lna.3}
\end{align}
where  $\mu_i=|\omega_i|+\kappa$ is the effective noise strength for oscillator $i$, $\eta_i$ is Gaussian white noise with $\langle \eta_i(t) \rangle = 0$ and $\langle \eta_i(t) \eta_j(t') \rangle = \delta_{ij}\delta(t-t')$, and where we have used the definition of the two global quantities
\begin{align}
	r \mathrm{e}^{\mathrm{i}\psi} = \frac{1}{N} \sum_j \mathrm{e}^{\mathrm{i}\Phi_j} \ , \qquad
	\tilde{r} \mathrm{e}^{\mathrm{i}\tilde{\psi}} = \frac{1}{N} \sum_j \xi_j \mathrm{e}^{\mathrm{i}\Phi_j} \ . \Bigg. \label{eq.lna.ops}
\end{align}
The first quantity is the Kuramoto order parameter associated with the `macroscopic' phases $\Phi_i$ and the second one convolves the macroscopic phases with the random variables $\xi_i$.

\begin{figure*}[t]
\begin{center}
\includegraphics[width=13.5cm]{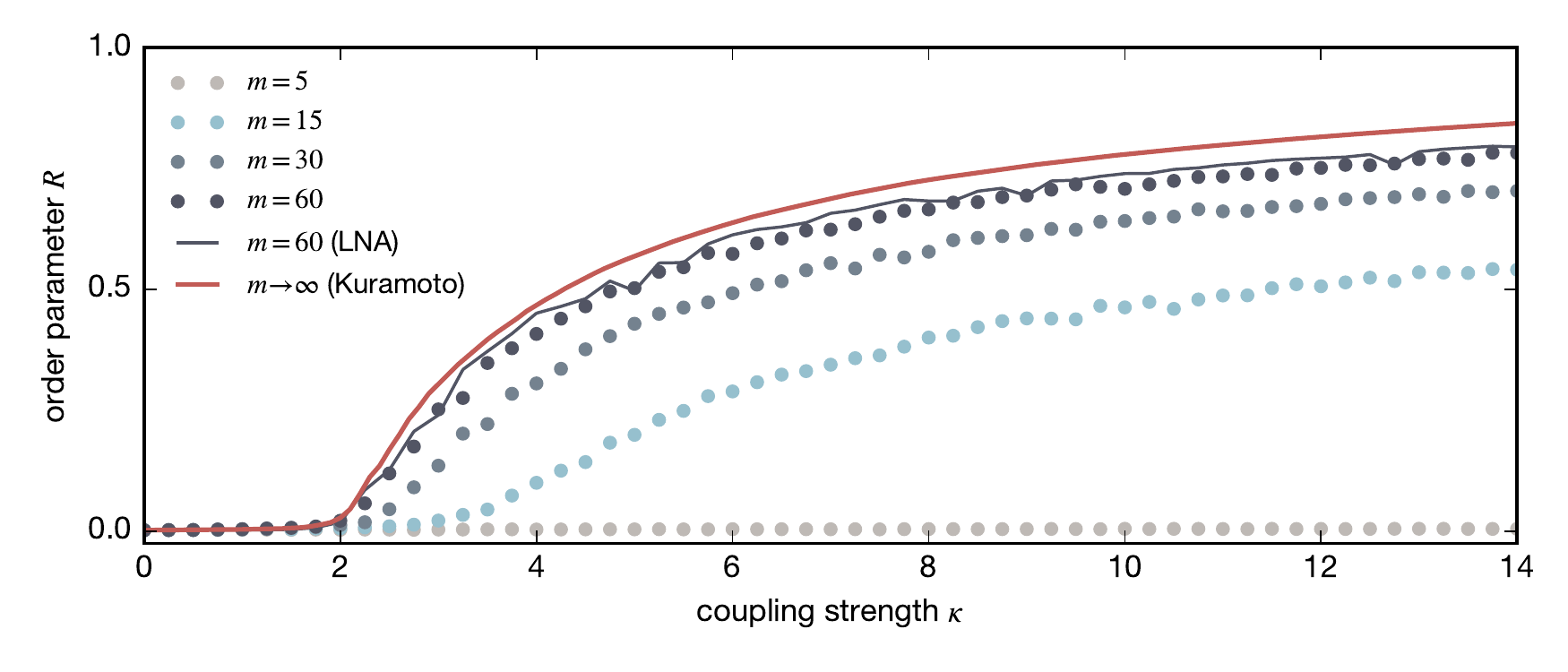}

\caption{Synchronization transition with the coupling strength as control parameter. The plot shows numerical results for the steady-state order parameter $R$ as a function of the coupling strength $\kappa$ for the model defined by Eqs.~(\ref{eq.gc.model}--\ref{eq.freq.dist}) for different phase discretizations $m$ (colored dots) as well as for the linear noise approximation (LNA) given by Eqs.~(\ref{eq.lna.2}) and (\ref{eq.lna.3}) (blue curve) and the deterministic Kuramoto model Eq.~(\ref{eq.kuramoto}) (red curve). Simulations involve $N=500$ oscillators and averages are taken over 25 realizations of the frequency distribution.}
\label{fig.phase.transition}
\end{center}
\end{figure*}

\subsection{Synchronization transient}

\noindent%
As for the case of two coupled oscillators, we assess the synchronization transient and the steady-state phase coherence for the many-oscillator system.
To this end, we consider the case of homogeneous frequencies $\omega_i = \omega$.
Fig.~\ref{fig.many.oscillators}a illustrates the synchronization transient by showing the time-dependent order parameter $\rho(t) = \langle |\Psi(t)|^2\rangle$ for different numbers of oscillators and phase discretizations (cf.~Fig.~\ref{fig.sync}a).
Figs.~\ref{fig.many.oscillators}b,c show the synchronization time $\tau_\mathrm{h}$, Eq.~(\ref{eq.sync.time.abs}), to reach an order parameter of $\rho=1/2$ from complete desynchronization as well as the time $\tau_\nu$, Eq.~(\ref{eq.sync.time.rel}), to reach a fraction $\nu=0.95$ of the steady-state order parameter $R$. Note that for coarse phase discretizations, the system may not reach an order parameter of $1/2$ at all, in which case the time $\tau_\mathrm{h}$ is undefined (Fig.~\ref{fig.many.oscillators}b).
Generally, the shown synchronization times, which characterize the nonlinear transient from complete desynchronization to synchrony, increase with the number of oscillators~$N$ but can exhibit a nonmonotonic behavior in the phase discretization $m$ for a large enough number of oscillators.

\subsection{Steady-state phase coherence and oscillator precision}

\noindent%
Turning to the steady-state phase coherence,
Fig.~\ref{fig.many.oscillators}d shows the steady-state order parameter $R$ as a function of the phase discretization for different numbers of oscillators (dots) as well as a comparison with the linear noise approximation (curves), Eqs.~(\ref{eq.lna.2}--\ref{eq.lna.ops}).
While the order parameter increases monotonically in the phase discretization, the behavior for larger numbers of oscillators hinting at a synchronization transition for a finite value of the phase discretization (dark dataset and arrowhead in Fig.~\ref{fig.many.oscillators}d).
This transition is likely related to the synchronization transition of the classical Kuramoto model in the presence of noise, where partial synchrony is enabled when the noise strength drops below a critical value that depends on the coupling strength \cite{Acebron2005}.
However, while in our case, the phase discretization is clearly related to the effective noise strength, cf.~Eq.~(\ref{eq.langevin}), it also introduces other effects such as sampling of the coupling function at discrete readout points, which may alter the behavior of the system apart from introducing fluctuations.

In the spirit of Section~\ref{subsec:precision.2osc}, we next assess the quality factor as the dimensionless ratio of the oscillation time scale and the exponential decay rate of the autocorrelation function.
Fig.~\ref{fig.many.oscillators}e shows the steady-state quality factor $Q$ computed from the average autocorrelation from all oscillators.
We find a massive increase in oscillator precision when the phase discretization reaches values for which also the onset of partial synchrony is observed, cf.~Fig.~\ref{fig.many.oscillators}d.
As expected, the quality factor also increases with the number of oscillators, a behavior that is well-known for coupled phase oscillator systems in general \cite{Cross2012}.
Figs.~\ref{fig.many.oscillators}d,e also suggest that for both the steady-state order parameter as well as the quality factor, the LNA specified by Eqs.~(\ref{eq.lna.2}--\ref{eq.lna.ops}) provides an excellent approximation in the limit of fine phase discretizations.

\subsection{Onset of synchronization for inhomogeneous frequencies}

\noindent%
Finally, we illustrate the behavior of many phase-discrete oscillators with inhomogeneous frequencies in the all-to-all coupled system Eq.~(\ref{eq.gc.model}).
To this end, we consider an ensemble of systems with quenched disorder, i.e., with intrinsic transition frequencies $\omega_i$ drawn from a specific distribution $f(\omega)$ but fixed for each realization of the system.
This scenario is well-studied for the classical Kuramoto model with unimodal and symmetric distributions $f$: In the thermodynamic limit of infinitely many oscillators, $N\to\infty$, this system exhibits a second-order synchronization phase transition at the critical coupling strength $\kappa_\mathrm{c} = 2/(\pi f(0))$ \cite{KuramotoBook1984,Strogatz2000}.
We here draw the transition frequencies $\omega_i$ from a Cauchy distribution centered around zero,
\begin{align}
	f(\omega) = \frac{1}{\pi} \frac{1}{1+\omega^2} \ ,
	\label{eq.freq.dist}
\end{align}
so that for the classical Kuramoto model in the thermodynamic limit, the phase transition occurs at $\kappa_\mathrm{c} = 2$.
Fig.~\ref{fig.phase.transition} shows the order parameter $R$
as a function of the coupling strength $\kappa$ for different phase discretizations $m$ and the Kuramoto limit $m\to\infty$.
Phase discretization decreases the limiting amount of synchrony and in some cases, even completely prohibits partial synchronization where the classical model is able to partially synchronize (see $m=5$ curve in Fig.~\ref{fig.phase.transition}).
Again, the LNA given by Eqs.~(\ref{eq.lna.2}--\ref{eq.lna.ops}) provides good agreement with the phase-discretized model in the limit of fine discretizations.

\section{Discussion}
\label{sec:discussion}

\noindent In this paper, we have presented a stochastic generalization of the Kuramoto model with discretized phases and investigated its synchronization behavior as well as its frequency fluctuations.
Remarkably, while the phase-discretized model converges towards the deterministic Kuramoto dynamics in the limit of a continuous phase, many key observables exhibit a non-monotonic behavior.
This leads to optima in the steady-state synchrony and oscillator precision for finite phase discretizations, which can exceed the corresponding values of the deterministic Kuramoto model.
These features arise from an interplay of different effects that are a consequence of the phase discretization such as discrete sampling of the coupling function and the inherent stochasticity of the coupling process.

The discretization schemes introduced here enable a straightforward implementation of coupled stochastic oscillations in a Markov chain setting and can be useful in coupling cyclic dynamics to mesoscopic systems. Such systems might include, e.g., chemical reaction networks \cite{Gillespie1977} and stochastic models of cell fate dynamics \cite{Klein2008}, where cyclic processes may effectively depict  periodic extrinsic signals such as the cell cycle \cite{Weber2014}, circadian rhythms \cite{Zwicker2010,Brown2014}, or periodic signaling activity \cite{Sugimoto2012}.
Moreover, it is straightforward to computationally generalize the phase-discretized model to the case of non-Markovian transitions between phase states that entail non-exponential waiting time distributions~\cite{Boguna2014}.

Here we have only taken a glimpse at the phenomena that can arise when phase-discretization of Kuramoto oscillators is combined with stochastic dynamics.
To illustrate their behavior, we have chosen the generic cases of two oscillators and many oscillators with all-to-all coupling; therefore, we could not address the spatiotemporal dynamics of spatially distributed systems such as those with short-range (e.g., nearest-neighbor) interactions, which may give rise to interesting patterning phenomena \cite{Escaff2014}.
Moreover, we have chosen a generic but contingent discretization scheme for the coupling process (see Appendix~\ref{app:alternative.couplings}).
It will be interesting to unfold the dynamics of different model realizations and to apply the proposed discretization schemes to, e.g., Kuramoto oscillators with inertia \cite{Tanaka1997,Olmi2014,Jorg2015} and excitable dynamics \cite{Lindner2004} as well as time-delayed coupling \cite{Schuster1989,Yeung1999,Earl2003,Jorg2014} and signal filtering \cite{Pollakis2014,Wetzel2017}, which goes beyond the Markovian approach.

\begin{acknowledgments}

\noindent%
I thank B.~D.~Simons for discussions and L.~Wetzel, L.~G.~Morelli, and I.~M.~Lengyel for critical comments on the paper.
I acknowledge the support of the Wellcome Trust (grant number 098357/Z/12/Z).

\end{acknowledgments}

\begin{appendix}

\section{Linear noise approximation of the phase-discretized model with coupling}
\label{app:linear.noise}

\noindent%
In this Appendix, we derive the linear noise approximation Eqs.~(\ref{eq.lna.phase}--\ref{eq.phase.diffusion}).
To this end, we perform a system size expansion of the system specified by Eqs.~(\ref{eq.me.main}--\ref{eq.cpl.operator}) in the standard way \cite{vanKampen2011}. The phase discretization $m$ is a natural candidate for the system size $\Omega$ as large $m$ lead to a more continuous phase.
For the oscillator system with phases $\boldsymbol{\varphi}=(\varphi_1,\hdots,\varphi_N)$, we define the `macroscopic' phases ${\boldsymbol{\Phi}}$ (that follow deterministic dynamics) and the random components ${\boldsymbol{\xi}}$ through the relation ${\boldsymbol{\varphi}} = \Omega {\boldsymbol{\Phi}} + \smash{\sqrt{\Omega}}{\boldsymbol{\xi}}$ where $\Omega=\varepsilon^{-1}=m/2\pi$. Furthermore, we define the probability distribution ${W}$ for the random components as ${W}(\boldsymbol{\xi},t)=P(\boldsymbol{\varphi}(\boldsymbol{\xi}),t)$. The next steps consist in calculating the time evolution of ${W}$, expanding in powers of $\sqrt{\Omega}$ and comparing coefficients.
The coefficients of $\sqrt{\Omega}$ yield the equation
\begin{align}
	\sum_i \frac{\partial {W}}{\partial \xi_i} \frac{\mathrm{d}\Phi_i}{\mathrm{d}t} 
	&= \sum_{i} \bigg\{ \omega_i  + \kappa \sum_{j} c_{ij}   \Gamma(\Phi_j-\Phi_i)   \bigg\}  \frac{\partial {W}}{\partial\xi_i} \ ,\label{eq.ln.macro}
\end{align}
whereas the coefficients of $\Omega^0$ result in
\begin{align}
\begin{split}
	\frac{\partial {W}}{\partial t}
	&=  \sum_{i} \frac {\partial }{\partial \xi _{i}}  \bigg\{   {\frac {|\omega_i |+ \kappa \sum_j c_{ij}}{2}}\frac {\partial {W}}{\partial \xi _{i}} \\
	&\qquad  -  \kappa \sum_{j} c_{ij} \Gamma'(\Phi_j-\Phi_i) (\xi_j-\xi_i)  {W} \bigg\}  \ ,
\end{split}\label{eq.ln.noise}
\end{align}
where $\Gamma'$ is the derivative of the coupling function.
Eq.~(\ref{eq.ln.macro}) describes the deterministic evolution of the macroscopic phases $\Phi_i$, while Eq.~(\ref{eq.ln.noise}) is a Fokker--Planck equation for the random components $\xi_i$.
The correspondence between Fokker--Planck and Langevin stochastic differential equations \cite{vanKampen2011} enables to immediately write Eqs.~(\ref{eq.lna.phase}--\ref{eq.phase.diffusion}) from Eqs.~(\ref{eq.ln.macro}) and (\ref{eq.ln.noise}).
In the case of no coupling, $\kappa=0$, Eqs.~(\ref{eq.lna.det}) and (\ref{eq.lna.noise}) reduce to $\mathrm{d}\Phi_i/\mathrm{d}t=\omega_i$ and $\mathrm{d}\xi_i/\mathrm{d}t = \smash{\sqrt{|\omega_i |}}\eta_i(t)$, so that the linear noise approximation Eq.~(\ref{eq.langevin}) derived for Eq.~({\ref{eq.single.osc}}) follows from Eqs.~(\ref{eq.lna.phase}--\ref{eq.phase.diffusion}) and $\omega_i \geq 0$.

\section{Alternative generalizations of coupling}
\label{app:alternative.couplings}

\noindent%
In this Appendix, we schematically discuss different possibilities to generalize the coupling term in a phase-discretized setting. To this end, we consider the Kuramoto model Eq.~(\ref{eq.kuramoto}) for the case of two coupled oscillators without self-coupling and $\omega_1=\omega_2=0$. 
Schematically, the time evolution of oscillator $i=1,2$ can now be written as $\mathrm{d}\phi_i/\mathrm{d}t={K}_i$, where ${K}_i$ represents the dynamical frequency contribution from its coupling term. For simplicity, here we neither address the dependence of the ${K}_i$ on the phases nor their implicit time-dependence; these do not add any qualitative features to our considerations.
We now illustrate different possibilities to generalize the coupling term by considering different stochastic processes (denoted by A, B, and C) for two discrete random variables $\varphi_1$ and $\varphi_2$ which all have in common that their expectation values satisfy $\mathrm{d}\langle \varphi_i \rangle/\mathrm{d}t = {K}_i$.

For case A, we introduce four non-negative rates $k_1^+$, $k_1^-$, $k_2^+$, and $k_2^-$, with the only constraint that they satisfy $k_i^+-k_i^-={K}_i$. The stochastic dynamics is defined by the master equation
\begin{align}
	\frac{\partial P_\mathrm{A}}{\partial t} &= \sum_{i=1}^2 \sum_{\sigma=\pm}  k_i^\sigma (\hat \varphi_i^{-\sigma} -1)  P_\mathrm{A} \ ,  \label{eq.p1}
\end{align}
where $P_\mathrm{A}=P_\mathrm{A}(\varphi_1,\varphi_2,t)$ and the $\hat \varphi_i^\pm$ are ladder operators, as defined in Eq.~(\ref{eq.ladder}).
Eq.~(\ref{eq.p1}) describes a system in which the forward and backward processes $\varphi \to \varphi+1$ and  $\varphi \to \varphi-1$ are independent for each oscillator, leading to four parallel processes with rates $k_i^\pm$. In this case, stochastic reactions do not conserve the total number $\varphi_1+\varphi_2$.
The coupling type investigated in this paper, Eq.~(\ref{eq.cpl.operator}), follows this spirit.

The stochastic dynamics of case B is defined by
\begin{align}
	\frac{\partial P_\mathrm{B}}{\partial t} &=\sum_{i=1}^2 |{K}_i|\sum_{\sigma=\pm}   \Theta(\sigma {K}_i) (\hat\varphi_i^{-\sigma}-1)  P_\mathrm{B} \ , \label{eq.p2}
\end{align}
Eq.~(\ref{eq.p2}) describes a process in which for each oscillator at each point in time, depending on the sign of ${K}_i$ either the process $\varphi \to \varphi+1$ or $\varphi \to \varphi-1$ can occur, as indicated by the Heaviside function $\Theta$.
A coupling in this spirit only admits a positive or negative frequency contribution at each point in time and importantly has zero contribution to the stochastic dynamics if ${K}_i=0$.
This is not the case for coupling type A, where $K_i=0$ only imposes $k_i^+=k_i^-$.

Case C is only possible if ${K}_1 = {K} = -{K}_2$; this is the case, e.g., for symmetric bidirectional coupling $c_{ij}=c_{ji}$ and an odd coupling function such as $\Gamma(\phi)=\sin\phi$. Here we introduce two non-negative rates $k^+$ and $k^-$ with the only constraint that they satisfy $k^+-k^-={K}$ and define the stochastic dynamics by
\begin{align}
	\frac{\partial P_\mathrm{C}}{\partial t} &= \sum_\sigma  k^\sigma (\hat \varphi_1^{-\sigma} \hat \varphi_2^\sigma -1)  P_\mathrm{C} \ , \label{eq.p3}
\end{align}
Eq.~(\ref{eq.p3}) describes a process in which the forward process  $\varphi \to \varphi+1$ in one oscillator is always accompanied by a backward process $\varphi \to \varphi-1$ in the other oscillator, leading to the `exchange of phase quanta' between the two oscillators and strict conservation of the total number $\varphi_1+\varphi_2$.
It is clear that such a coupling type only works for symmetric coupling as any coupling-induced reaction will affect both oscillators involved.

This list of generalizations is by no means exhaustive and only gives a flavor of the different types of implementations of the stochastic coupling process.
For instance, additional possibilities arise from the differences in how oscillators might internally process the coupling signals from different oscillators, e.g., whether they are processed independently \cite{Wetzel2017} or first integrated and then processed as a whole \cite{Pollakis2014}.
The adequate formalization to describe a specific system depends on the physical implementation of the coupling process.

\section{Steady-state order parameter and cross correlation of the two-oscillator system}
\label{appendix:exact.signal}

\noindent In this Appendix, we derive Eq.~(\ref{eq.ss}) for the steady-state expectation value of the order parameter for two coupled phase-discretized oscillators.
First, we obtain a master equation for the discrete phase difference $\theta=\varphi_1-\varphi_2$ by using Eq.~(\ref{eq.me.main}) for $N=2$ and $c_{ij}=1-\delta_{ij}$ and marginalizing over one of the discrete phase variables, $\tilde{P}(\theta,t) = \sum_{\varphi_1} P(\varphi_1,\varphi_1-\theta,t)$. For simplicity, we only consider the case $\omega_1 \geq 0$, $\omega_2 \geq 0$; the other cases follow analogously. Hence, we obtain the master equation for $\tilde{P}$ as
\begin{align}
\begin{split}
	\varepsilon \frac{\partial \tilde{P}}{\partial t} &= \bigg\{ \omega_1 (\hat\theta^{-} - 1)  + \omega_2 (\hat\theta^{+} - 1) \\
	&\qquad + 2\kappa \bigg( \sum_{\sigma=\pm} [ \tilde\Gamma (\varepsilon\theta +\sigma \varepsilon) ]_\sigma \hat\theta^{\sigma} - 1 \bigg) \bigg\} \tilde{P} ,
\end{split} \label{eq.prob.difference}
\end{align}
where $\hat\theta^\pm$ are the ladder operators for the phase difference and where we have used the same convention for $[\cdot]_\pm$ as in Eq.~(\ref{eq.gc.cpl}).
Here, $\tilde\Gamma(\phi)=(\Gamma(\phi)-\Gamma(-\phi))/2$ is the odd part of the coupling function and
for the Kuramoto--Sakaguchi-type coupling Eq.~(\ref{eq.KS.coupling}), this evaluates to $\smash{\tilde\Gamma}(\phi)=\cos(\phi_0)\sin(\phi)$.
Next, we define the steady-state expectation values ${X}_n =\langle \mathrm{e}^{\mathrm{i}n\varepsilon\theta} \rangle$ and using the master equation~(\ref{eq.prob.difference}), we obtain their time evolution  as
\begin{align}
\begin{split}
	\varepsilon \frac{\mathrm{d}{X}_n}{\mathrm{d}t} 
	&= \varepsilon \sum_\theta \frac{\partial \tilde{P}(\theta,t)}{\partial t} \mathrm{e}^{\mathrm{i}n\varepsilon\theta} \\
	&=  [(\mathrm{e}^{\mathrm{i}n\varepsilon}-1)\omega_1 +(\mathrm{e}^{-\mathrm{i}n\varepsilon}-1)\omega_2 \\
	&\qquad + 2\kappa(\cos(n \varepsilon)-1)]{X}_n \\
	&\qquad - \kappa \cos(\phi_0) \sin(n\varepsilon)({X}_{n+1}-{X}_{n-1}) \ . 
\end{split} \label{eq.fourier.hierarchy}
\end{align}
The key observation is that from the definition of the ${X}_n$ and $\varepsilon=2\pi/m$, it follows that ${X}_0 = 1$ and ${X}_m=1$, so that the set of equations given by (\ref{eq.fourier.hierarchy}) constitutes a closed hierarchy for the ${X}_n$ with $0 \leq n \leq m$.
At steady state, $\mathrm{d} {X}_n/\mathrm{d}t=0$, this yields the following set of algebraic equations,
\begin{align}
\begin{split}
	{X}_0 &= 1 \ , \\
	\lambda_n {X}_n &= {X}_{n-1}-{X}_{n+1} \ , \qquad (1 \leq n \leq m-1) \\
	{X}_m &= 1 \ ,
\end{split} \label{eq.hierarchy}
\end{align}
where the $\lambda_n$ are defined in Eq.~(\ref{eq.eta}).
Solving this hierarchy starting from $n=m-1$, each expectation value ${X}_n$ can be expressed in terms of the next lower expectation value ${X}_{n-1}$. It can be shown by induction that this leads to the generic form
\begin{align}
	{X}_n = \prod_{i=n}^{m-1} \Lambda_i-\Lambda_n {X}_{n-1} \ , \quad  (1 \leq n \leq m-1) \label{eq.hierarchy.2}
\end{align}
where the $\Lambda_n$ satisfy the nonlinear recurrence relation
\begin{align}
	\Lambda_n = \frac{1}{\Lambda_{n+1}-\lambda_n} \ , \label{eq.recurrence.relation}
\end{align}
with initial condition $\Lambda_m=0$.
The continued fraction Eq.~(\ref{eq.contfrac}) is the solution to this recurrence relation as is obvious from repeatedly inserting Eq.~(\ref{eq.recurrence.relation}) into itself.
Since the cross correlation is given by ${X} = {X}_1$, its exact solution is obtained from Eq.~(\ref{eq.hierarchy.2}) as
\begin{align}
	X = \prod_{i=1}^{m-1} \Lambda_i-\Lambda_1 \ ,  \label{eq.cc.ss}
\end{align}
and Eq.~(\ref{eq.ss}) for the order parameter follows from this via Eq.~(\ref{eq.order}) as $R=(1+\operatorname{Re}X)/2$. 

Since the solution given by Eq.~(\ref{eq.cc.ss}) is somewhat opaque due to its combinatorial complexity, we here give explicit expressions for ${X}$ for small $m$,
\begin{align*}
	{X}\big|_{m=2} &= 0 \ , \
	&{X}\big|_{m=3} &= \frac{\sqrt{3} {p} + {q} -1}{3 {p} ^2- {q} ^2-1} \ , \bigg. \\[2pt]
	{X}\big|_{m=4} &= \frac{1}{{p} - {q} } \ , \
	&{X}\big|_{m=6} &= \bigg[ \frac{1}{\sqrt{3} {p} -{q} }+\frac{{p}-\sqrt{3}q}{\sqrt{3}} \bigg]^{-1} \ , \\
\end{align*}
where
\begin{align*}
	{p} = \frac{\omega_1+\omega_2 + 2\kappa}{\kappa\cos\phi_0} \ , \qquad
	{q} = \mathrm{i}\frac{\omega_1-\omega_2}{\kappa\cos\phi_0} \ .
\end{align*}

\end{appendix}

\end{document}